\documentclass[floats,prb,aps]{revtex4}
\usepackage{graphicx}
\begin{document}


\title{Collective properties of magnetobiexcitons in quantum wells' and graphene superlattices}

\author{Oleg L. Berman$^{1}$,  Roman Ya. Kezerashvili$^{1}$ and Yurii E. Lozovik$^{2}$}

\affiliation{\mbox{$^{1}$Physics Department, New York City College of Technology, the City University of New York,}
\\ Brooklyn, NY 11201, USA
 \\ \mbox{$^{2}$ Institute of Spectroscopy, Russian Academy of
Sciences,}  \\ 142190 Troitsk, Moscow Region, Russia}


\begin{abstract}
We propose the Bose-Einstein condensation (BEC) and superfluidity of  quasi-two-dimensional (2D)
spatially indirect magnetobiexcitons in  a slab of superlattice with
alternating electron and hole layers  consisting from the semiconducting quantum wells (QWs) and graphene superlattice
in high magnetic field. The two different Hamiltonians of
a dilute gas of magnetoexcitons with a dipole-dipole repulsion in superlattices consisting of both QWs and graphene layers in the limit of high magnetic field
have been reduced  to one effective Hamiltonian a dilute gas of
two-dimensional excitons without magnetic field.  Moreover, for $N$ excitons we have reduced the problem of $2N\times 2$ dimensional space onto the problem of $N\times 2$  dimensional space by integrating over the coordinates of the relative motion of an electron (e) and a hole (h). The instability of the ground state of the system of interacting
two-dimensional indirect magnetoexcitons in a slab of superlattice with
alternating electron and hole layers  in high magnetic field is established. The stable
system of  indirect quasi-two-dimensional magnetobiexcitons, consisting
of pair of indirect excitons with opposite dipole moments is
considered. The density of superfluid
component $n_{s}(T)$ and
 the temperature of the Kosterlitz-Thouless phase transition to the superfluid state in
the  system of  two-dimensional indirect magnetobiexcitons, interacting as
electrical quadrupoles, are obtained for both QW and graphene realizations.

\vspace{0.3cm}

PACS numbers:  71.35.Ji, 71.35.Lk, 71.35.-y, 68.65.Cd, 73.21.Cd

\end{abstract}

\maketitle


\section{Introduction}
\label{int}

The many-particle systems of the spatially-indirect excitons  in
coupled quantum wells (CQWs) in the presence or absence of a
magnetic field $B$ have been the subject of recent experimental
investigations \cite{Snoke,Butov,Timofeev,Eisenstein}. These systems
are of interest, in particular, in connection with the possibility
of Bose-Einstein condensation (BEC) and superfluidity of indirect
excitons or electron-hole pairs, which would manifest itself in the
CQW as persistent electrical currents in each well and also through
coherent optical properties and Josephson
phenomena.\cite{Lozovik,Birman,Littlewood,Vignale,Berman} In high
magnetic fields, two-dimensional (2D) excitons survive in a
substantially wider temperature range, as the exciton binding
energies increase with magnetic
field.\cite{Lerner,Paquet,Kallin,Yoshioka,Ruvinsky,Ulloa,Moskalenko}
 The problem of essential interest is also collective properties of magnetoexcitons in high magnetic fields
 in  superlattices and layered system.\cite{Filin}

 In this Paper we propose a new physical realization of magnetoexcitonic BEC
and superfluidity in in superlattices with
alternating electronic and hole  layers, that is in a sense
representing array of CQWs or graphene layers with spatially separated  electrons
($e$) and holes $h$ with spatially separated electrons and holes in high magnetic field.  Recent technological advances have allowed the production of graphene,
 which is a 2D honeycomb lattice of carbon atoms that form the basic planar structure in graphite \cite{Novoselov1,Zhang1} Graphene has been attracting a great deal
of experimental and theoretical attention because of unusual
properties in its band structure
\cite{Novoselov2,Zhang2,Nomura,Jain}. It is a gapless semiconductor
with massless electrons and holes which have been described as
Dirac-fermions \cite{DasSarma}.   Since there is no gap between the
conduction and valence bands in graphene without magnetic
 field, the screening effects result in the absence of excitons in
graphene in the absence of  magnetic field.  A strong magnetic field
produces a gap since the energy spectrum becomes discrete  formed by
Landau levels. The gap reduces screening  and leads to the formation
of magnetoexcitons. We also consider magnetoexcitons in the
superlattices with  alternating electronic and hole  graphene
layers. We suppose that recombination times can be much greater
 than relaxation times $\tau _{r}$
due to small  overlapping of spatially separation of e- and h- wave
functions in CQW or graphene layers. In this case electrons and
holes are characterized by different quasi-equilibrium chemical
potentials.
 Then  in the system of indirect excitons
in superlattices, as in CQW \cite{Lozovik,Berman}, the
quasiequilibrium phases appear \cite{remark}. While coupled-well
structures with spatially separated electrons and holes are
typically considered to be under applied electrical field, which
separates electrons and holes in different quantum
wells\cite{Butov,Timofeev}, we assume there are no external fields
applied to a slab of superlattice.  If "electron" and "hole" quantum
wells alternate, there are excitons with parallel dipole moments in
one pair of wells, but dipole moments of excitons in another
neighboring pairs of neighboring wells have opposite direction. This
fact leads to essential distinction of properties of  $e-h$ system
in superlattices from ones for coupled quantum wells with spatially
separated electrons and holes (where indirect exciton system is
stable due to dipole-dipole repelling of all excitons). This
difference manifests itself already beginning from three-layer
$e-h-e$ or $h-e-h$ system. We assume that alternating $e-h-e$ layers
can be formed by independent gating with the corresponding
potentials which shift chemical potentials in neighboring layers up
and down or by alternating doping (by donors and acceptors,
respectively).

In this Paper we reduce the
problem of magnetoexcitons to the problem of
  excitons at $B = 0$.  The unstability of the ground state of the
system of interacting indirect excitons in  slab of superlattice
with alternating $e-$ and $h-$ layers  is established in strong magnetic field.
Two-dimensional indirect  magnetobiexcitons, consisting of
 the indirect magnetoexcitons with
opposite dipole moments, are considered in high magnetic field. The
radius and the binding energy  of  indirect magnetobiexciton are
calculated. These magnetobiexcitons repel each other as electrical
quadrupoles at long distances. In result, the system of indirect
magnetobiexcitons is stable. In the ladder approximation collective
spectrum of the weakly interacting by  the quadrupole law
two-dimensional indirect magnetobiexcitons is considered. The
superfluid density $n_{s}(T)$ of  interacting two-dimensional
indirect magnetobiexcitons  in  superlattices is calculated at low
temperatures $T$. We analyze the dependence of Kosterlitz-Thouless
transition\cite{Kosterlitz} temperature
  and superfluid density on magnetic field.

The rest of the Paper is organized  as   follows.  In Sec.\  \ref{effect_Ham}, we derive an effective Hamiltonian of
magnetoexcitons in both CQWs and two graphene layers in strong magnetic field.  In Sec.\  \ref{instab}, we prove the instability of dipole magnetoexcitons
in QW and graphene superlattices due to the attraction of oppositely directed dipoles. BEC and superfluidity of quadrupole magnetobiexcitons in QW and graphene
 superlattices has been established and analyzed  in Sec.\  \ref{bec_b}.
We present and discuss the numerical results   in Sec.\  \ref{disc}.

\section{Effective Hamiltonian of magnetoexcitons in strong magnetic field}
\label{effect_Ham}

We start with two interacting excitons in CQW in the presence of the
external magnetic field $\mathbf{B}$. Without loss of generality we
can take the magnetic field to be in the \textit{z} direction. We
find it convenient to work with
the symmetric gauge for a vector potential $\mathbf{A}_{e(h)}=1/2[\mathbf{B}%
\times \mathbf{r}_{e(h)}]$. We employ Jacobi coordinates for the exciton
center of mass $\mathbf{R}=(m_{e}\mathbf{r}_{e}+m_{h}\mathbf{r}%
_{h})/(m_{e}+m_{h})$ and relative motion of electrons ($e$) and holes ($h$) $%
\mathbf{r}=\mathbf{r}_{e}-\mathbf{r}_{h}$ and in these coordinates the
Hamiltonian $\hat{H}$ for 2D spatially separated excitons is
\begin{eqnarray}\label{hamil}
&&\hat{H}=\int d\mathbf{R}\int d\mathbf{r}\left[ \hat{\psi}^{\dagger }(%
\mathbf{R},\mathbf{r})\sum_{i=e,h}\left( \frac{1}{2m_{i}}\left(
\mathbf{p}_{i}+\frac{e}{c}\mathbf{A}_{i}\right) ^{2}\right. \right.
\nonumber  \label{H_Tot} \\
&-&\left. \left. \frac{e^{2}}{\epsilon \sqrt{(\mathbf{r}_{e}-\mathbf{r}%
_{h})^{2}+D^{2}}}\right) \hat{\psi}(\mathbf{R},\mathbf{r})\right]   \nonumber
\\
&+&\frac{1}{2}\int d\mathbf{R}_{1}\int d\mathbf{r}_{1}\int d\mathbf{R}%
_{2}\int d\mathbf{r}_{2}\hat{\psi}^{\dagger }(\mathbf{R}_{1},\mathbf{r}_{1})%
\hat{\psi}^{\dagger }(\mathbf{R}_{2},\mathbf{r}_{2})  \nonumber \\
&&\sum_{i,j=e,h}U^{ij}(\mathbf{r}_{i1}-\mathbf{r}_{j2})\hat{\psi}(%
\mathbf{R}_{2},\mathbf{r}_{2})\hat{\psi}(\mathbf{R}_{1},\mathbf{r}_{1}),
\end{eqnarray}%
%
where $\hat{\psi}^{\dagger }(\mathbf{R},\mathbf{r})$ and $\hat{\psi}(\mathbf{%
R},\mathbf{r})$ are the creation and annihilation operators for
magnetoexcitons; $\mathbf{r}_{e}$ and $\mathbf{r}_{h}$ are electron
and hole locations along quantum wells, respectively; $D$ is the
distance between electron and hole quantum wells, $e$ is the charge
of an electron; $c$ is the speed of light and $\epsilon $ is a
dielectric constant (for graphene layers $\epsilon $ is the
dielectric constant of the matrix in which graphene layers are
embedded). In Eq.~(\ref{hamil}) we use the two-particle potentials
$U^{ij}$ for the electron-electron, hole-hole, electron-hole and
hole-electron interaction, for electrons and
holes from two different excitons: $U^{ee}(\mathbf{r}_{e1}-\mathbf{r}%
_{e2})=e^{2}/(\epsilon |\mathbf{r}_{e1}-\mathbf{r}_{e2}|)$, $U^{hh}(\mathbf{r%
}_{h1}-\mathbf{r}_{h2})=e^{2}/(\epsilon |\mathbf{r}_{h1}-\mathbf{r}_{h2}|)$,
$U^{eh}(\mathbf{r}_{e1}-\mathbf{r}_{h2})=-e^{2}/(\epsilon \sqrt{|\mathbf{r}%
_{e1}-\mathbf{r}_{h2}|^{2}+D^{2}})$, $U^{he}(\mathbf{r}_{h1}-\mathbf{r}%
_{e2})=-e^{2}/(\epsilon \sqrt{|\mathbf{r}_{h1}-\mathbf{r}_{e2}|^{2}+D^{2}})$.

The four-component Hamiltonian of an isolated electron-hole  pair  in bilayer graphene
with spatially separated electrons ($e$) and holes ($h$) in one valley in magnetic field $B$ neglecting the Coulomb interaction is given by\cite{Iyengar}
\begin{eqnarray}\label{Ham}
\hat{H}= v_{F}\left(\begin{array}{cccc}
0 &p_{x}^{(e)}+ i p_{y}^{(e)}&0&0 \\
p_{x}^{(e)} - i p_{y}^{(e)}&0&0&0 \\
0&0&0& p_{x}^{(h)} - i p_{y}^{(h)}\\
0 &0& p_{x}^{(h)}+ i p_{y}^{(h)}&0 \\
\end{array}   \right),
\end{eqnarray}
where
\begin{eqnarray}\label{peh}
\mathbf{p}^{(e)} = -i\hbar\nabla_{e}+
\frac{e}{c}\mathbf{A}_{e}; \hspace{1cm}
\mathbf{p}^{(h)} = -i\hbar\nabla_{h}-
\frac{e}{c}\mathbf{A}_{h};
\end{eqnarray}
where $e$ is an electron charge; $c$ is the speed of light; $\mathbf{r}_{e}$ and $\mathbf{r}_{h}$ are two-dimensional (2D) vectors of coordinates
 of an electron and hole, correspondingly;
$\mathbf{A}_{e}$ and $\mathbf{A}_{h}$ are the vector potential of an
electron and hole, correspondingly; $v_{F} = \sqrt{3}at/(2\hbar)$ is
the Fermi velocity of electrons in graphene ($a =2.566 \AA$  is the
lattice constant, $t \approx 2.71 eV$ is the overlap integral
between the nearest carbon atoms)\cite{Lukose}.

Note that the ratio of the contribution to the energy from the Zeeman term $%
\Delta E_{Z}(B)$ to the characteristic separation between the nearest Landau
levels $\Delta E_{L}(B) $ is negligible (for quantum wells: this ratio is $%
\Delta E_{Z}(B)/\Delta E_{L}(B) \approx m_{e}/m_{e}^{(0)} \approx 0.1$ ; for
graphene layers: at $B=10 T$ this ratio is $\Delta E_{Z}(B)/\Delta E_{L}(B)
\approx \mu_{B}B/(\sqrt{2}\hbar v_{F}r_{B}^{-1}) \approx 5\times 10^{-3}$,
where $\mu_{B} = \hbar e/(2m_{e}^{(0)}c)$ is the Bohr magneton; $m_{e}
\approx 0.1 m_{e}^{(0)}$ is the effective electron mass in $GaAs/AlGaAs$
quantum wells; $m_{e}^{(0)}$ is the mass of a free electron). Therefore, the
contributions to the single-electron Hamiltonian from the Zeeman splitting
 set identically to zero analogously to Refs.~[%
\onlinecite{Jain,Iyengar}]. We take into account the energy
degeneracy corresponding to two possible spin projections in quantum
wells and graphene and two graphene valleys (two pseudospins). Since
electrons on a graphene lattice can be in two valleys, there are
four types of excitons in bilayer graphene. Due to the fact that all
these types of excitons have identical envelope wave functions and
energies\cite{Iyengar}, we consider below only excitons in one
valley. Also, we use $n_{0} = n/(4s)$ as the density of excitons in
graphene superlattice, with $n$ denoting the total density of
excitons, $s$ is the spin degeneracy (equal to $4$ for
magnetoexcitons in bilayer graphene). Besides, we use $n_{0} = n/s$
as the density of excitons in quantum wells, with $n$ denoting the
total density of excitons, $s$ is the spin degeneracy (equal to $4$
for magnetoexcitons in coupled quantum wells).

A conserved quantity for an isolated electron-hole pair in magnetic field $B$
is the exciton generalized momentum $\hat{\mathbf{P}}$ defined as
\[
\hat{\mathbf{P}}=-i\hbar \nabla _{e}-i\hbar \nabla _{h}+\frac{e}{c}(\mathbf{A%
}_{e}-\mathbf{A}_{h})-\frac{e}{c}[\mathbf{B}\times (\mathbf{r}_{e}-\mathbf{r}%
_{h})]\
\]%
for the Dirac equation in graphene layers\cite{Iyengar} as well as for the
Schr\"{o}dinger equation in CQWs\cite{Gorkov,Lerner,Kallin}.

The Hamiltonian of a single isolated magnetoexciton without any random field
($V_{e}(\mathbf{r}_{e})=V_{h}(\mathbf{r}_{h})=0$) is commutated with $\hat{%
\mathbf{P}}$, and hence  they have the same the eigenfunctions, which have
the following form (see Refs.~[\onlinecite{Lerner,Gorkov}]):
\begin{eqnarray}
&&\Psi _{k\mathbf{P}}(\mathbf{R},\mathbf{r})=  \nonumber  \label{W_Func_Gen}
\\
&&\exp \left\{ i\mathbf{R}\left( \mathbf{P}+\frac{e}{2}\mathbf{B}\times
\mathbf{R}\right) +i\gamma \frac{\mathbf{P}\mathbf{r}}{2}\right\} \Phi _{k}(%
\mathbf{P},\mathbf{r}),
\end{eqnarray}%
where $\Phi _{k}(\mathbf{P},\mathbf{r})$ is a function of internal
coordinates $\mathbf{r}$ and the eigenvalue $\mathbf{P}$ of the generalized
momentum, and $k$ represents the quantum numbers of exciton internal motion.
The wave function of the relative coordinate for $e$ and $h$ spatially separated in different graphene layers $\tilde{\Phi}(\mathbf{r})$ can be expressed in terms of the two-dimensional harmonic oscillator eigenfunctions $\Phi_{n_{1},n_{2}}(\mathbf{r})$.
For an electron in Landau level $n_{+}$ and a hole in level $n_{-}$, the four-component wave functions for the relative coordinate are\cite{Iyengar}
\begin{eqnarray}\label{electron}
\tilde{\Phi}_{n_{+},n_{-}}(\mathbf{r}) =   \left( \sqrt{2}\right)^{\delta_{n_{+},0}+\delta_{n_{-},0}-2}\left( \begin{array}{c} s_{+}s_{-}
\Phi_{|n_{+}|-1,|n_{-}|-1}(\mathbf{r})\\
s_{+}\Phi_{|n_{+}|-1,|n_{-}|}(\mathbf{r})\\
s_{-}\Phi_{|n_{+}|,|n_{-}|-1}(\mathbf{r})\\
\Phi_{|n_{+}|,|n_{-}|}(\mathbf{r})
\end{array}\right) ,
\end{eqnarray}
where $s_{\pm} = \mathrm{sgn} (n_{\pm})$. The corresponding energy of the electron-hole pair $E_{n_{+},n_{-}}^{(0)}$ (which is the eigenvalue of the Hamiltonian (Eq.~(\ref{Ham}))) is given by\cite{Iyengar}
\begin{eqnarray}\label{energy2}
E_{n_{+},n_{-}}^{(0)} =  \frac{\hbar v_{F}}{r_{B}} \sqrt{2}\left[\mathrm{sgn}(n_{+})\sqrt{|n_{+}|} - \mathrm{sgn}(n_{-})\sqrt{|n_{-}|}\right].
\end{eqnarray}
where $r_{B} = \sqrt{c\hbar/(eB)}$ is a magnetic length; $v_{F} = \sqrt{3}at/(2\hbar)$
is the Fermi velocity of electrons in graphene ($a =2.566 \AA$  is
a lattice constant, $t \approx 2.71 eV$ is the overlap integral
between the nearest carbon atoms).\cite{Lukose} The two-dimensional harmonic oscillator wave functions eigenfunctions corresponding to $e$ and $h$ in spatially separated CQWs $\Phi_{n_{1},n_{2}}(\mathbf{r})$ are given by\cite{Iyengar}
\begin{eqnarray}\label{electron_S}
 \Phi_{n_{1},n_{2}}(\mathbf{r}) &=& (2\pi)^{-1/2}2^{-|m|/2}\frac{\tilde{n}!}{\sqrt{n_{1}!n_{2}!}} \frac{1}{r_{B}} \mathrm{sgn}(m)^{m}\frac{r^{|m|}}{r_{B}^{|m|}} \nonumber \\
&& \exp\left[-im\phi - \frac{r^{2}}{4r_{B}^{2}}\right] L_{\tilde{n}}^{|m|}\left(\frac{r^{2}}{2r_{B}^{2}}\right) ,
\end{eqnarray}
where $L_{\tilde{n}}^{|m|}$  denotes Laguerre polynomials; $m = n_{1} -n_{2}$; $\tilde{n} = \min(n_{1},n_{2})$, and $\mathrm{sgn}(m)^{m}\rightarrow 1$ for $m=0$.
The expression for $\Phi _{k}(\mathbf{P},\mathbf{r})$ for CQW is given in
Ref.~[\onlinecite{Lerner,Ruvinsky}], and in Ref.~[\onlinecite{Iyengar}]) for
graphene layers. In high magnetic fields the magnetoexcitonic quantum
numbers $k=\{n_{+},n_{-}\}$ for an electron in Landau level $n_{+}$ and a
hole in level $n_{-}$; $\gamma =(m_{h}-m_{e})/(m_{h}+m_{e})$.

In a strong magnetic field at low densities (~$n\ll r_{B}^{-2}$~) ($r_{B}=%
\sqrt{\hbar c/(eB)}$ is the magnetic length) indirect magnetoexcitons repel
as parallel dipoles, and we have for the pair interaction potential:
\[
\hat{U}(|\mathbf{R}_{1}-\mathbf{R}_{2}|)\equiv \hat{U}^{ee}+\hat{U}^{hh}+%
\hat{U}^{eh}+\hat{U}^{he}\simeq \frac{e^{2}D^{2}}{\epsilon |\mathbf{R}_{1}-%
\mathbf{R}_{2}|^{3}}.
\]%
If we expand the magnetoexciton field operators in a single magnetoexciton
basis set $\Psi _{k\mathbf{P}}(\mathbf{R},\mathbf{r})$: $\hat{\psi}^{\dagger
}(\mathbf{R},\mathbf{r})=\sum_{k\mathbf{P}}\Psi _{k\mathbf{P}}^{\ast }(%
\mathbf{R},\mathbf{r})\hat{a}_{k\mathbf{P}}^{\dagger }$; $\hat{\psi}(\mathbf{%
R},\mathbf{r})=\sum_{k\mathbf{P}}\Psi _{k\mathbf{P}}(\mathbf{R},\mathbf{r})%
\hat{a}_{k\mathbf{P}}$, where $\hat{a}_{k\mathbf{P}}^{\dagger }$ and $\hat{a}%
_{k\mathbf{P}}$ are the corresponding creation and annihilation operators of
a magnetoexciton in $(k,\mathbf{P})$ space and substitute the expansions for
the field creation and annihilation operators into the Eq.~(\ref{H_Tot}) and
obtain the effective Hamiltonian in terms of creation and annihilation
operators in $\mathbf{P}$ space. In high magnetic field, one can ignore
transitions between Landau levels and consider only the lowest Landau level
states (for CQWs $n_{+}=n_{-}=0$; for graphene layers $n_{+}=n_{-}=1$).

\bigskip

\bigskip

Using the orthonormality of the functions $\Phi _{k}(\mathbf{0},\mathbf{r})$
we obtain the effective Hamiltonian $\hat{H}_{\mathrm{eff}}$ in strong
magnetic fields.

Due to the orthonormality of the wave functions $\Phi _{n_{+},n_{-}}(\mathbf{%
0,r})$ the projection of the Hamiltonian (Eq.~(\ref{H_Tot})) onto the lowest
Landau level results in the effective Hamiltonian, which does not reflect
the spinor nature of the four-component magnetoexcitonic wave functions in
graphene. Since typically, the value of $r$ is $r_{B}$, and $P\ll \hbar
/r_{B}$ in this approximation, the effective Hamiltonian $\hat{H}_{\mathrm{%
eff}}$ in the magnetic momentum representation $P$ in the subspace the
lowest Landau level (for QWs $n_{+}=n_{-}=0$; for graphene layers $%
n_{+}=n_{-}=1$) has the same form (compare with Ref.[\onlinecite{Berman}])
as for two-dimensional boson system without a magnetic field, but with the
magnetoexciton magnetic mass $m_{B}$ (which depends on $B$ and $D$; see
below) instead of the exciton mass ($M=m_{e}+m_{h}$), magnetic momenta
instead of ordinary momenta and renormalized random field (for the lowest
Landau level we denote the spectrum of the single exciton $\varepsilon
_{0}(P)\equiv \varepsilon _{00}(\mathbf{P})$):
\begin{eqnarray}
&&\hat{H}_{\mathrm{eff}}=\sum_{\mathbf{P}}\varepsilon _{0}(P)\hat{a}_{%
\mathbf{P}}^{\dagger }\hat{a}_{\mathbf{P}}+\frac{1}{2}  \nonumber
\label{H_eff} \\
&&\times \sum_{\mathbf{P}_{1},\mathbf{P}_{2},\mathbf{P}_{3},\mathbf{P}%
_{4}}\left\langle \mathbf{P}_{1},\mathbf{P}_{2}\left\vert \hat{U}\right\vert
\mathbf{P}_{3},\mathbf{P}_{4}\right\rangle \hat{a}_{\mathbf{P}_{1}}^{\dagger
}\hat{a}_{\mathbf{P}_{2}}^{\dagger }\hat{a}_{\mathbf{P}_{3}}\hat{a}_{\mathbf{%
P}_{4}},
\end{eqnarray}%
where the matrix element $\left\langle \mathbf{P}_{1},\mathbf{P}%
_{2}\left\vert \hat{U}\right\vert \mathbf{P}_{3},\mathbf{P}_{4}\right\rangle
$ is the Fourier transform of the pair interaction potential $U(R)=e^{2}D^{2}/\epsilon R^{3}$. For an isolated magnetoexciton on the lowest Landau level at the small
magnetic momenta under consideration, $\varepsilon _{0}(\mathbf{P})\approx
P^{2}/(2m_{B})$, where $m_{B}$ is the effective \textit{magnetic} mass of a
magnetoexciton in the lowest Landau level and is a function of the distance $%
D$ between $e$ -- and $h$ -- layers and magnetic field $B$ (see Ref.[%
\onlinecite{Ruvinsky}]). In strong magnetic fields at $D\gg r_{B}$ the
exciton magnetic mass is $m_{B}(D)=\epsilon D^{3}/(e^{2}r_{B}^{4})$ for QWs
\cite{Ruvinsky} and $m_{B}(D)=\epsilon D^{3}/(4e^{2}r_{B}^{4})$ for graphene
layers \cite{Berman_Lozovik_Gumbs}.

Note that the ratio of the contribution to the energy from the Zeeman term $\Delta E_{Z}(B)$ to the characteristic separation between the nearest Landau levels $\Delta E_{L}(B) $ is negligible (for quantum wells: this ratio is $\Delta E_{Z}(B)/\Delta E_{L}(B) \approx m_{e}/m_{e}^{(0)} \approx 0.1$ ; for graphene layers: at $B=10 T$ this ratio is $\Delta E_{Z}(B)/\Delta E_{L}(B) \approx \mu_{B}B/(\sqrt{2}\hbar v_{F}r_{B}^{-1}) \approx 5\times 10^{-3}$, where $\mu_{B} = \hbar e/(2m_{e}^{(0)}c)$ is the Bohr magneton; $m_{e} \approx 0.1 m_{e}^{(0)}$ is the effective electron mass in $GaAs/AlGaAs$ quantum wells; $m_{e}^{(0)}$ is the mass of a free electron). Therefore, the contributions to the single-electron Hamiltonian  from the Zeeman splitting and very small pseudospin splitting in graphene (caused by two valleys in graphene) set identically to zero analogously to Refs.~[\onlinecite{Jain,Iyengar}]. We assume the energy degeneracy respect to two possible spin projections in quantum wells and graphene and two graphene valleys  (two pseudospins).
Since electrons on a  graphene lattice can be in two
valleys, there are four types of excitons in bilayer
graphene. Due to the fact that all these types of excitons have identical
envelope wave functions and energies\cite{Iyengar}, we consider below only
excitons in one valley.
Also, we use $n_{0} = n/(4s)$ as the density of excitons
in graphene superlattice,  with $n$ denoting the total density of excitons, $s$ is the spin degeneracy (equal to $4$ for magnetoexcitons in bilayer graphene).
Besides, we use $n_{0} = n/s$ as the density of excitons
in quantum wells,  with $n$ denoting the total density of exciton, $s$ is the spin degeneracy (equal to $4$ for magnetoexcitons in coupled quantum wells).

For large electron-hole separation $D\gg r_B$,
transitions between  Landau levels due to the Coulomb electron-hole
attraction can be neglected, if the following condition is valid, (for QWs $E_{b} = e^{2}/(\epsilon_b D) \ll \hbar \omega_{c} = \hbar e B(m_{e} + m_{h})/(2m_{e}m_{h}c) $; for graphene layers $E_{b} = 4 e^{2}/(\epsilon D) \ll \hbar v_{F}/r_B$; where $E_{b}$ and $\omega_{c}$ are the magnetoexcitonic binding energy and the cyclotron frequency, correspondingly).
This corresponds to high magnetic field $B$, large interlayer
separation $D$ and large dielectric constant of the insulator layer
between the graphene layers. In this notation, $v_{F} = \sqrt{3}at/(2\hbar)$
is the Fermi velocity of electrons. Also, $a =2.566 \AA$  is a lattice
constant, $t \approx 2.71 eV$ is the overlap integral between
nearest carbon atoms \cite{Lukose}.

\section{Instability of dipole magnetoexcitons in QW and graphene superlattices}
\label{instab}

Let us show the low-density system of weakly interacting
two-dimensional indirect magnetoexcitons in superlattices is instable,
contrary to two-layer system in CQW. At small densities  $nr_{B}^2 \ll
1$ the system of indirect excitons at low temperatures is the
two-dimensional weakly nonideal Bose
 gas with normal to wells dipole moments  ${\bf d}$ in the ground state
 ($d \sim eD$, $D$ is the interwell separation),
increasing with the distance between wells $D$.
 In contrast to ordinary excitons,
for  low-density spatially indirect magnetoexciton system the main
contribution to the energy is originated from dipole-dipole
 interactions  $U_{-}$ and $U_{+}$ of magnetoexcitons with opposite and
parallel  dipoles, respectively. Two parallel ($+$) and opposite
($-$) dipoles in low-density system interact as $U_{+} = - U_{-} =
e^{2}D^{2}/\epsilon R^{3}$, where $\epsilon $ is the
dielectrical constant; $R$ is the distance between dipoles along
wells planes; we suppose that $D/R \ll 1$ and $L/R \ll 1$ ($L$ is
the mean distance between dipoles normal to the wells). We
consider the case, when the number of quantum wells $k$ in
superlattice is restricted $k \ll \frac{1}{D\sqrt{\pi n}}$, and
this is valid for small $k$ or for sufficiently low exciton
density ($n$ is exciton surface density).

The distinction between magnetoexcitons and bosons manifests itself in
exchange effects.\cite{Berman,Snoke_book}
The exchange interaction in  spatially separated system
is  suppressed in contrast to  $e-h$ system in one well
due to smallness of tunnel exponent $T$ connected with the penetration
through barrier of dipole-dipole interaction.
Hence, at $D \gg r{B}$  exchange phenomena,
connected with the distinction between excitons and bosons,
can be neglected for the both QWs and graphene layers\cite{Berman_Lozovik_Gumbs}.

For the analysis of stability of the ground state of the weakly
nonideal Bose gas of indirect excitons in superlattices we apply the
Bogolubov approximation. The total Hamiltonian  $\hat H_{tot}$ of
the low-density system of indirect excitons in superlattice is given by:$
\hat H_{tot} = \hat H_{0} + \hat H_{int}$.
Here  $\hat H_{0}$ is the effective Hamiltonian of the
system of noninteracting magnetoexcitons:
\begin{eqnarray}
\label{h0}
\hat H_{0} = \sum_{{\bf p}}^{}\varepsilon _{0}(p)
(a_{{\bf p}}^{+}a_{{\bf p}}^{} + b_{{\bf p}}^{+}b_{{\bf p}}^{}
+ a_{-{\bf p}}^{+}a_{-{\bf p}}^{} + b_{-{\bf p}}^{+}b_{-{\bf p}}^{})  ,
\end{eqnarray}
where $\varepsilon _{0}(p) = p^{2}/(2m_{B})$ is the spectrum of
isolated two-dimensional indirect magnetoexciton; $\mathbf{p}$ represents the excitonic magnetic momentum.
$a_{{\bf p}}^{+}$, $a_{{\bf p}}^{}$, $b_{{\bf p}}^{+}$, $b_{{\bf p}}^{}$
are creation and annihilation operators of magnetoexcitons
with up and down dipoles;
$\hat H_{int}$ is the effective Hamiltonian of the interaction
between magnetoexcitons:
\begin{equation}
\begin{array}{c}
\label{hint}
\hat H_{int} = \frac{U}{2S}
\sum_{{\bf p}_{1} + {\bf p}_{2} = {\bf p}_{3} + {\bf p}_{4}}^{}
 (a_{{\bf p}_{4}}^{+}a_{{\bf p}_{3}}^{+}a_{{\bf p}_{2}}a_{{\bf p}_{1}}  +
\nonumber  \\
 b_{{\bf p}_{4}}^{+}b_{{\bf p}_{3}}^{+}b_{{\bf p}_{2}}b_{{\bf p}_{1}}
- a_{{\bf p}_{4}}^{+}a_{{\bf p}_{3}}^{+}b_{{\bf p}_{2}}b_{{\bf p}_{1}}
- b_{{\bf p}_{4}}^{+}b_{{\bf p}_{3}}^{+}a_{{\bf p}_{2}}a_{{\bf p}_{1}} -
\nonumber \\
 a_{{\bf p}_{4}}^{+}b_{{\bf p}_{3}}^{+}a_{{\bf p}_{2}}b_{{\bf p}_{1}})  ,
\end{array}
\end{equation}
$S$ is the surface of the system. Let us consider the temperature $T
= 0$. Assuming the majority of particles are in the condensate ($N -
N_{0}/N_{0} \ll 1$, where $N$ and $N_{0}$ are the total number of
particles and the  number of particles in condensate), we account as
in Bogolubov approximation only interaction between condensate
particles and excited particles with  condensate particles,
neglecting by the interaction between noncondensate particles. Then
the total Hamiltonian transforms to:
\begin{equation}
\label{htot1}
\begin{array}{c}
\hat H_{tot} = \frac{1}{2}\sum_{{\bf p} \ne 0}^{}  [\varepsilon _{0}(p)
(a_{{\bf p}}^{+}a_{{\bf p}}^{} + b_{{\bf p}}^{+}b_{{\bf p}}^{}
+ a_{-{\bf p}}^{+}a_{-{\bf p}}^{} +
 \nonumber  \\
b_{-{\bf p}}^{+}b_{-{\bf p}}^{})
 -  Un(a_{{\bf p}}^{+}b_{-{\bf p}}^{+} + a_{{\bf p}}^{}b_{-{\bf p}}^{} +
  a_{-{\bf p}}^{+}b_{{\bf p}}^{+} + a_{-{\bf p}}^{}b_{{\bf p}}^{}
   \nonumber   \\
+ a_{{\bf p}}^{+}b_{{\bf p}}^{} + a_{-{\bf p}}^{+}b_{-{\bf p}}^{}
+ a_{{\bf p}}^{}b_{{\bf p}}^{+} + a_{-{\bf p}}^{}b_{-{\bf p}}^{+})] ,
\end{array}
\end{equation}
In Eq.(\ref{htot1}) terms, arising from
first and second terms of the Hamiltonian Eq.(\ref{hint}),
which describe the repulsion
of the indirect magnetoexcitons with parallel dipole moments, are compensating
by other terms of the Hamiltonian Eq.(\ref{hint}), describing
the attraction  of indirect magnetoexcitons with opposite dipoles.
In the result only terms describing the attraction survive.
Let us diagonalize  Hamiltonian
 $\hat H_{tot}$ by using of the unitary transformation
of the Bogolubov type\cite{Abrikosov}
\begin{equation}
\begin{array}{c}
\label{bog}
a_{\bf p} = \frac{1}{\sqrt{1 - A_{{\bf p}}^{2} - B_{{\bf p}}^{2}}}
 (\alpha _{\bf p} + A_{\bf p}\alpha _{-{\bf p}}^{+} +
 B_{\bf p}\beta _{-{\bf p}}^{+})  ;  \nonumber \\
b_{\bf p} = \frac{1}{\sqrt{1 - A_{{\bf p}}^{2} - B_{{\bf p}}^{2}}}
 (\beta _{\bf p} + A_{\bf p}\beta _{-{\bf p}}^{+} +
B_{\bf p}\alpha _{-{\bf p}}^{+})  ,
\end{array}
\end{equation}
where the coefficients $A_{\bf p}$ and $B_{\bf p}$ are found
from the condition of vanishing of coefficients at
nondiagonal terms in Hamiltonian.
In result we obtain
\begin{eqnarray}
\label{hamquas}
\hat H_{tot} = \sum_{{\bf p} \ne 0}^{}  \varepsilon (p)
(\alpha _{{\bf p}}^{+}\alpha _{{\bf p}}^{} +
\beta _{{\bf p}}^{+}\beta _{{\bf p}}^{})
\end{eqnarray}
with  the  spectrum of
quasiparticles $\varepsilon (p)$:
\begin{eqnarray}
\label{quasisp}
\varepsilon (p) = \sqrt{(\varepsilon_{0}(p))^{2} - (nU)^{2}}  .
\end{eqnarray}
At small momenta $p < \sqrt{2m_{B}nU}$
the spectrum of excitations becomes imaginary.
 Hence, the system of weakly interacting
 indirect magnetoexcitons in slab of superlattice
is unstable. It can be seen that the condition of the instability of magnetoexcitons as stronger as magnetic field higher, because $m_{B}$ increases with the increase of magnetic field, and, therefore, the region of $p$ resulting in the imaginary collective spectrum increases as $B$ increases.

This, on first view, strange result can be illustrated by the
following example. There are equal number of dipoles oriented up
and down. Let us consider four dipoles, two of them being oriented
up and two --- down. It is easy to count that number of repelling
pairs is smaller than that of attracting ones. The prevailing of
attraction leads to unstability.

\section{BEC and superfluidity of quadrupole magnetobiexcitons in QW and graphene superlattices}
\label{bec_b}

\begin{figure}
\includegraphics[width=1.9in]{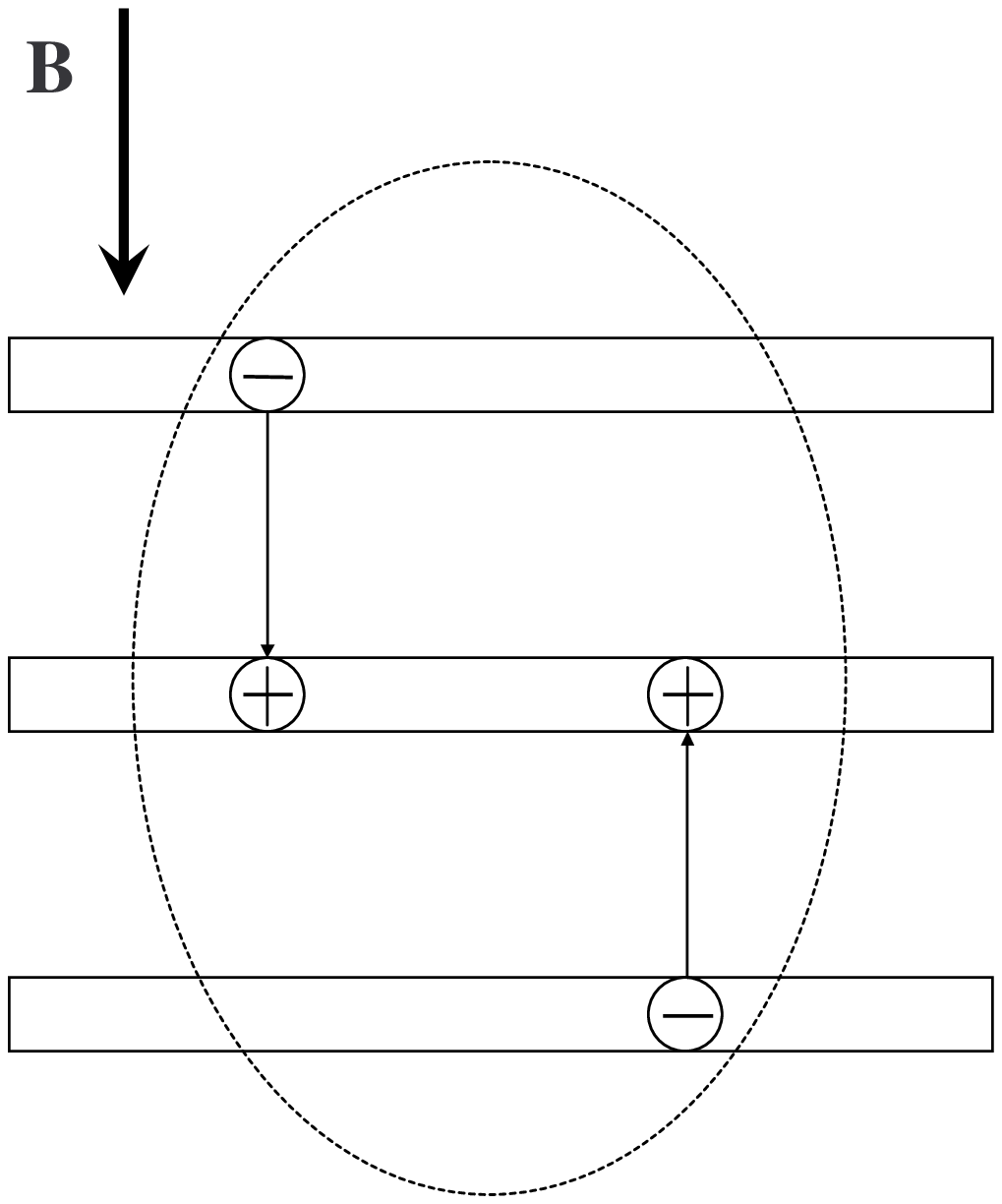}
\caption{Two-dimensional indirect magnetobiexcitons consisting of indirect magnetoexcitons with opposite dipole
moments, located in neighboring pairs of wells/graphene layers.}
\label{biexciton}
\end{figure}

\begin{figure}
\includegraphics[angle=-90,width = 2.2in]{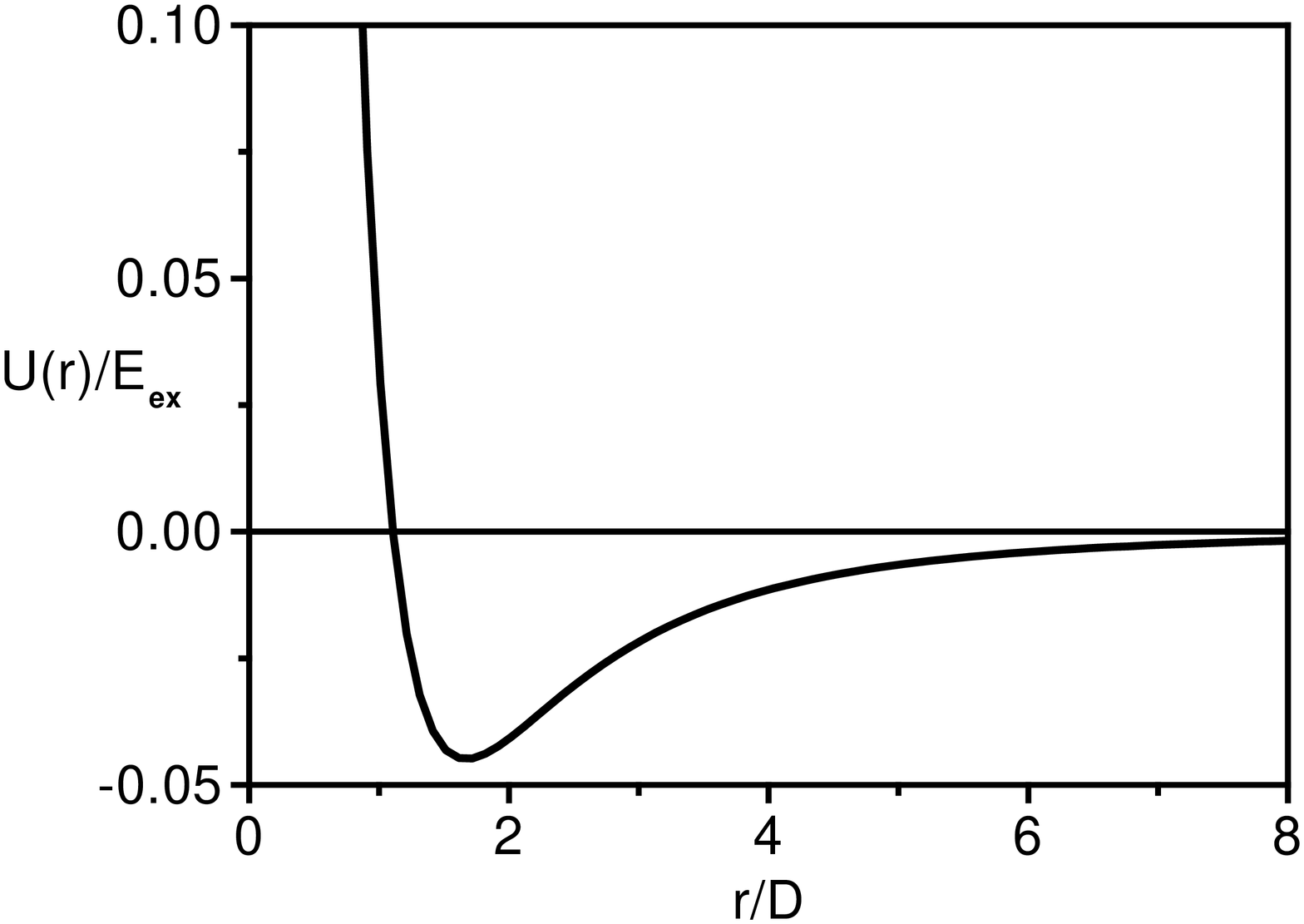}
\caption{The potential energy $U(r)$ of the interaction of
 indirect magnetoexcitons with opposite dipoles, located
 in neighboring pairs of wells or graphene layers (in units of the binding energy of
indirect magnetoexciton $E_{ex} = e^{2}/\epsilon D$), as a  function of
the distances $r$ between magnetoexcitons along the wells/graphene layers (in units of $D$).}
\label{interaction}
\end{figure}

Let us consider as the ground state of the system the low-density
weakly nonideal gas of two-dimensional indirect magnetobiexcitons,
created by indirect magnetoexcitons with opposite dipoles
 in neighboring pairs of wells (Fig.~\ref{biexciton}).
The small parameter at the adiabatic approximation is  the numerical
small parameter which is equal to the ratio of magnetobiexciton and
 magnetoexciton energies or the ratio between radii of magnetoexciton and
magnetobiexciton along quantum wells (or graphene layers) (see, e.g., Ref.~[\onlinecite{Musin}]). These
parameters are small, and they are even smaller
than analogous parameters for atoms and molecules. The smallness
of these parameters will be verified below by the results of the
calculation of indirect magnetobiexciton. Here it was assumed, that the
distance between wells (or graphene layers) $D$ is greater than the radius of indirect
magnetobiexciton $a_{b}$. The potential energy of
interaction between indirect magnetoexcitons with opposite dipoles $U(r)$
has the form shown on Fig.~\ref{interaction}  ($r$ is the distance between indirect
magnetoexcitons along quantum wells/graphene layers):

\begin{eqnarray}
\label{potbi}
U(r) = \frac{e^{2}}{\epsilon r}  -  \frac{2e^{2}}{\epsilon \sqrt{r^{2} +
D^{2}}}  + \frac{e^{2}}{\epsilon \sqrt{r^{2} + 4 D^{2}}}  .
\end{eqnarray}
 At $r > 1.11 D$
indirect magnetoexcitons attract, and at $r < 1.11 D$  they repel.
The minimum of potential energy
$U(r)$ locates at
 $r = r_{0} \approx 1.67 D$ between indirect excitons.
At large $D$ one can expand the potential energy $U(r)$ on
 the parameter $r - r_{0}/D \ll 1$:
\begin{eqnarray}
\label{riad}
U(r) = - 0.04 \frac{e^{2}}{\epsilon D} + 0.44 \frac{e^{2}}{\epsilon D^{3}}
(r - r_{0})^{2}  .
\end{eqnarray}
So at large $D$ magnetobiexciton levels correspond to the two-dimensional harmonic
oscillator with the frequency
$\omega = 0.88 e^{2}/(m_{B} \epsilon D^{3})$:
\begin{eqnarray}
\label{eqv}
E_{n} = - 0.04 \frac{e^{2}}{\epsilon D} +
 2\sqrt{2}E_{0}\left(\frac{r^{*}}{D}\right)^{3/2}(n + 1)    ,
\end{eqnarray}
where $E_{0} = m_{B}e'^{4}/(\hbar ^{2}\epsilon) $,
$r^{*} = \hbar ^{2}\epsilon/(2m_{B}e'^{2})$, $e'^{2} = 0.88 e^{2}$.
In the ground state the characteristic spread
 of magnetobiexciton $a_{b}$ along quantum wells/graphene layers
 (near the mean radius of magnetobiexciton
$r_{0}$ along wells/graphene layers) is:
\begin{eqnarray}
\label{rad}
a_{b} = \sqrt{\frac{2\hbar }{m_{B}\omega }} = (8r^{*})^{1/4}D^{3/4}
= 1.03 a_{ex}  ,
\end{eqnarray}
where $a_{ex} = (8r^{ex})^{1/4}D^{3/4}$;
$r^{ex} = \hbar ^{2}\epsilon/(2m_{B}e^{2})$ is the two-dimensional
effective Bohr radius with the effective magnetic mass $m_{B}$.
Hence,
the ratio of the binding energies of magnetobiexciton and magnetoexciton is:
$E_{bex}/E_{ex} = 0.04 \ll 1$  at
$D \gg a_{ex}$ (the
 ratio of radii of magnetoexciton and magnetobiexciton is
 $r_{0}/a_{ex} = 0.67 (8r^{ex})^{1/4}D^{-1/4} \ll 1$).
So adiabatic condition is valid.

The mean dipole moment of indirect
magnetobiexciton is equal to zero. However,
the quadrupole moment is nonzero and equal to $Q = 3eD^{2}$
(the large axis of the quadrupole is normal to quantum wells/graphene layers).
So indirect magnetobiexcitons interact at long distances $R \gg D$ as
parallel quadrupoles:  $U(R) = 9e^{2}D^{4}/(\epsilon R^{5})$.

Exchange effects,
 connected with the distinction between low-density indirect magnetobiexcitons and bosons,
can be suppressed due to the negligible overlapping of wave functions of two
magnetobiexcitons on account of the potential barrier, associated with the
quadrupole repulsion of indirect magnetobiexcitons  at long distances
analogously to dipole magnetoexcitons.
At large $D$ the  small tunnelling parameter connected with this barrier
has the form $exp[-0.93\sqrt{D/r^{ex}}]$.
Hence, at $D \gg r^{ex}$  exchange effects
for indirect magnetobiexcitons can be neglected.

We account the scattering of magnetobiexciton on magnetobiexciton
by using of the results of the theory of two-dimensional Bose-gas.\cite{Lozovik}
The chemical potential $\mu $ of two-dimensional biexcitons,
 repulsed by the quadrupole law, in the
 ladder approximation,   has the form
 (compare to Refs.~[\onlinecite{Lozovik,Berman}]):
 \begin{eqnarray}
\mu =  \frac{4\pi \hbar^{2}n_{bex}}{m_{B}^{b} \log \left[ \hbar^{4/3}
\epsilon^{2/3}/\left(8\pi (9 m_B^b e^2 D^4)^{2/3}n_{bex}\right) \right]} .
\label{Mu}
\end{eqnarray}
 where $n_{bex} = n/(2s)$
is the density of magnetobiexcitons in QWs and $n_{bex} = n/(8s)$ in graphene layers; $m_{B}^{b} = 2m_{B}$ is the mass of a magnetobiexciton.

At small momenta the collective spectrum of magnetobiexciton system
is the sound-like $\varepsilon (p) = c_{s}p$
  ($c_{s} = \sqrt{\mu/m_{B}^{b}}$ is
the sound velocity) and  satisfied
to Landau criterion for superfluidity.
The density of the superfluid component
 $n_S (T)$  for two-dimensional system
with the sound spectrum can be estimated as:\cite{Griffin}
\begin{equation}
\label{n_s}
n_S(T) = n_{bex} - \frac{3 \zeta (3) }{2 \pi \hbar^{2}}\frac{k_{B}^{3}T^3}{m_{B}^{b}c_s^4} .
\end{equation}
The second term in Eq.(\ref{n_s})  is the temperature dependent
 normal density  taking into account  gas
of phonons ("bogolons") with dispersion law $\varepsilon (p) =
\sqrt{\mu/m_{B}^{b}}p$, $\mu $ is given by Eq.(\ref{Mu}).

\begin{figure}
\includegraphics[width = 2.2in]{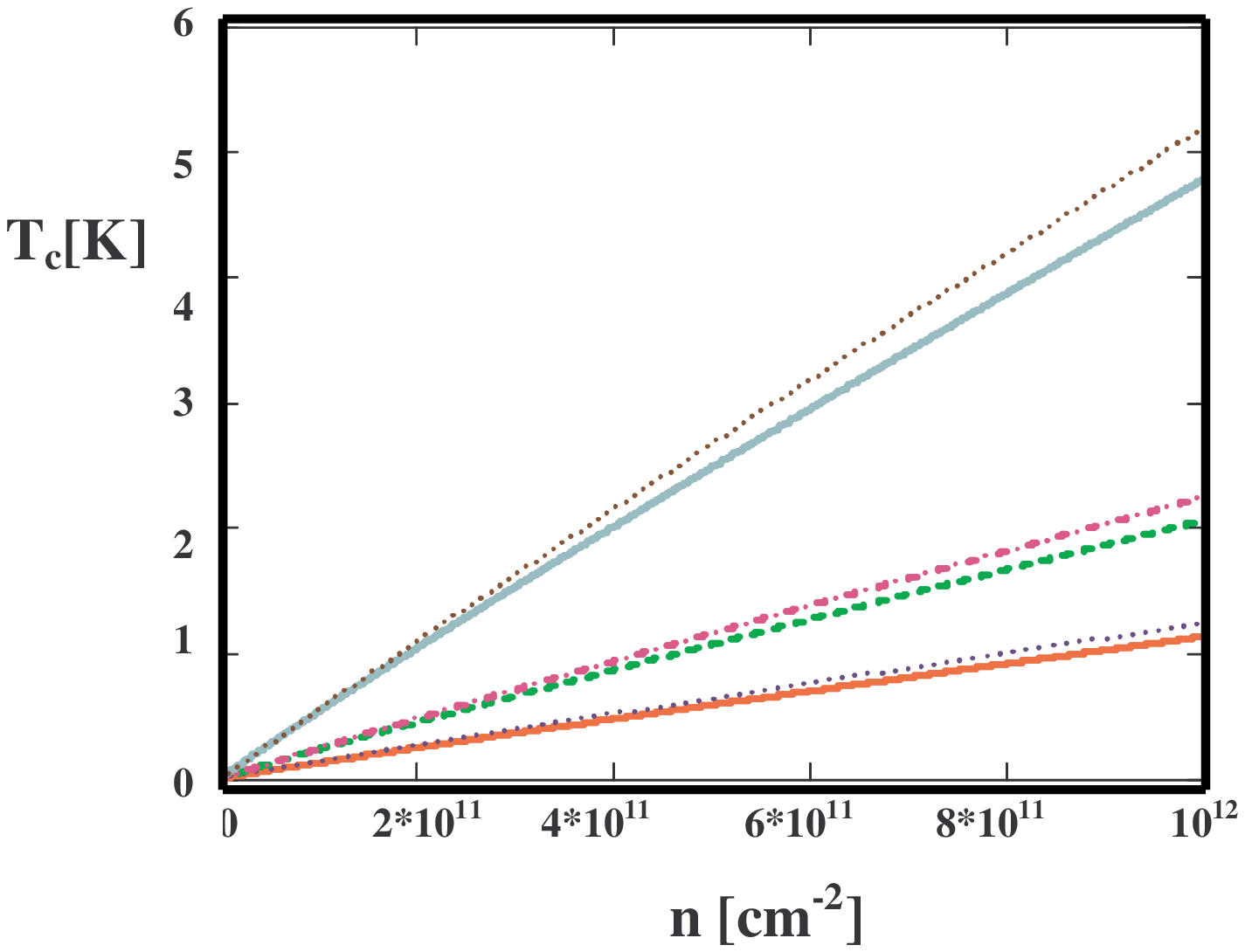}
\caption{Dependence of temperature of Kosterlitz-Thouless transition
$T_c = T_c (B)$ (in units $K$) for the superlattice consisting of quantum wells (for $GaAs/AlGaAs$: $\epsilon = 13$); and
(for graphene layers separated by the layer of $SiO_{2}$: $\epsilon = 4.5$)
on
the magnetoexciton density $n$ (in units $cm^{-2}$) at the interlayer
separation $D = 10 nm$  at
different magnetic fields $B$: $B = 20 T$ -- solid
curve for quantum wells and dotted curve for graphene layers; $B = 15 T$ --
dashed curve for quantum wells and dashed-dotted curve for graphene layers; $B = 10 T$ -- thin solid curve for quantum wells and solid and thin dotted curve for graphene layers.}
\label{THD}
\end{figure}

In a 2D system, superfluidity of magnetobiexcitons appears below the
Kosterlitz-Thouless transition temperature $T_{c} = \pi n_{S}(T)/(2m_{B}^{b})$, where only coupled vortices are
present \cite{Kosterlitz}. Employing $n_S(T)$ for
the superfluid component, we obtain an
equation for the Kosterlitz-Thouless transition temperature
$T_c$ with solution
\begin{eqnarray}
 && T_c = \left[\left( 1 +
\sqrt{\frac{32}{27}\left(\frac{m_B^{b}k_BT_c^0}{\pi \hbar^{2} n_{bex}}\right)^3 +
1} \right)^{1/3}  \right.
 \nonumber\\
 &&- \left. \left( \sqrt{\frac{32}{27}
\left(\frac{4 m_B^{b} k_BT_c^0}{\pi \hbar^{2} n_{bex}}\right)^3 + 1} - 1 \right)^{1/3}\right]
\frac{T_c^{0}}{ 2^{1/3}}\ .
\label{tct}
\end{eqnarray}
Here, $T_c^{0}$ is an auxiliary quantity, equal to the temperature
at which the superfluid density vanishes in the mean-field
approximation, i.e., $n_S(T_c^{0}) = 0$, $T_c^0 = k_B^{-1} \left(2
\pi \hbar^{2} n c_s^4 m_B^{b}/(3 \zeta (3))\right)^{1/3}$. The
temperature $T_c^0 = T_c^0(B,D)$ may be used to estimate the
crossover region where local superfluid density appears for
magnetobiexcitons  on a scale smaller or of the order of the mean
intervortex separation in the system. The local superfluid density
can manifest itself in local optical or transport properties. The
dependence of $T_c$ on the density of magnetoexcitons at different
magnetic field $B$ for superlattice consisting of quantum wells and
graphene layers is represented on Fig.~\ref{THD}.

\section{Discussion}
\label{disc}

It is shown that the low-density system of indirect magnetoexcitons
in a slab of superlattice  consisting of alternating $e$ and $h$ in
QWs or GLs in high magnetic field occur to be \textit{%
instable} due to the attraction of magnetoexcitons with opposite
dipoles at large distances. Note that in spite of both QW and
graphene realizations are represented by completely different
Hamiltonians, the effective Hamiltonian in a strong magnetic field
was obtained to be the same.  Moreover, for $N$ excitons we have
reduced the number of the degrees of freedom from $2N\times 2$ to
$N\times 2$  by integrating over the coordinates of the relative
motion of e and h. The instability of
 the ground state of the system of interacting
two-dimensional indirect magnetoexcitons in a slab of superlattice
with alternating electron and hole layers of both QWs and GLs in
high magnetic field is claimed due to the attraction between the
indirect excitons with opposite directed dipole moments. The stable
system occurs to be  indirect quasi-two-dimensional
magnetobiexcitons, consisting from indirect excitons with opposite
directed dipole moments. The stability of the system is connected
with the quadrupole-quadrupole repulsion of indirect
magnetoexcitons.  So at the pumping increase at low temperatures the
excitonic line must vanish and only magnetobiexcitonic line
survives.
 The Kosterlitz-Thouless transition to the superfluid
state is calculated for the system of indirect magnetobiexcitons.
According to Eq.\ (\ref{tct}),  the temperature $T_c$ for the onset
of superfluidity due to the Kosterlitz-Thouless transition at {\it a
fixed magnetobiexciton density} decreases as a function of magnetic
field $B$ and interlayer separation $D$. This is due to the
increased effective magnetic mass $m_B$ of magnetoexcitons as a
functions of $B$ and $D$. The $T_c$ decreases as $B^{-{1}/{2}}$ at
$D \ll r_B$ or as $B^{-2}$ when $D \gg r_B$.
 According to Fig.~\ref{THD}, the  Kosterlitz-Thouless temperature $T_{c}$ is higher for the superlattice consisting of
 graphene layers than for the superlattice consisting of the quantum wells, and this difference is as stronger as the magnetic field is smaller.

\acknowledgments
Yu~.E.~L. was supported by grants from RFBR and INTAS.

\end{document}